\documentclass[twocolumn,showpacs,preprintnumbers,amsmath,amssymb,prb,floatfix,superscriptaddress]{revtex4}

\usepackage[defblank]{paralist}
\usepackage{amsmath}
\usepackage{amssymb,theorem}
\usepackage{nicefrac}
\usepackage[dvips]{graphics}
\usepackage{mathrsfs}
\usepackage{color}
\usepackage{dcolumn}
\usepackage[mathcal]{euscript}
\usepackage{url}
\usepackage{graphicx,tabularx}
\usepackage{rotating}
\usepackage{appendix}

\begin{document}
\title{Probability of Inconsistencies in Theory Revision\\
\normalsize{A multi-agent model for updating logically interconnected beliefs under bounded confidence}}
\author{Sylvia Wenmackers}
\affiliation{University of Groningen, Faculty of Philosophy, Oude Boteringestraat 52, 9712 GL Groningen, The Netherlands}
\email{s.wenmackers@rug.nl}\thanks{Corresponding author}
\author{Danny E.~P.~Vanpoucke}
\affiliation{Ghent University, Department of Inorganic and Physical Chemistry, Krijgslaan 281 -- S3, 9000 Gent, Belgium}
\author{Igor Douven}
\affiliation{University of Groningen, Faculty of Philosophy, Oude Boteringestraat 52, 9712 GL Groningen, The Netherlands}


\begin{abstract}
We present a model for studying communities of epistemically interacting agents who update their belief states by averaging (in a specified way) the belief states of other agents in the community. The agents in our model have a rich belief state, involving multiple independent issues which are interrelated in such a way that they form a theory of the world. Our main goal is to calculate the probability for an agent to end up in an inconsistent belief state due to updating (in the given way). To that end, an analytical expression is given and evaluated numerically, both exactly and using statistical sampling. It is shown that, under the assumptions of our model, an agent always has a probability of less than 2\,\% of ending up in an inconsistent belief state. Moreover, this probability can be made arbitrarily small by increasing the number of independent issues the agents have to judge or by increasing the group size. A real-world situation to which this model applies is a group of experts participating in a Delphi-study.
\end{abstract}
\maketitle


\section{Introduction}\label{sec:intro}
Sociophysics studies social phenomena using existing models from statistical physics, such as the Ising spin model \cite{Sznajd:2000} or the concept of active Brownian particles \cite{Schweitzer:2003}, or tailor-made models that are physical in spirit, including various multi-agent models \cite{HegselmannKrause:2002}. The use of agent-based models has become increasingly popular in social dynamics research \cite{MacyWiller:2002,MacyFlache:2009}. The branch of sociophysics we are interested in here is opinion dynamics, which investigates groups of epistemically interacting agents. There is evidence from psychological experiments that agents adjust their opinion when they are informed about the opinion of another agent \cite{Wood:2000}. Partly inspired by such accounts, opinion dynamics studies how the opinions or belief states of agents evolve over time as a result of interactions with other agents, which may lead to cluster formation, polarization, or consensus on the macro-level \cite{HegselmannKrause:2002}. Opinion dynamics is also of interest for social epistemology---a branch of philosophy, that focuses on social aspects of knowledge of beliefs \cite{Goldman:1999}. It studies, for example, how to deal with peer disagreement, the best response to which is found to be context-sensitive; among other factors, it depends on the goal of the investigation \cite{DouvenKelp:2011}.

In general, the processes involved in opinion dynamics are very complex. Apart from analytical results, also computer simulations of agent-based models are used to study these large, complex systems. Simulations allow researchers to perform pseudo-experiments in situations in which real-life experiments are impossible, impractical, or unethical to perform. For some seminal contributions to the field of computational opinion dynamics, see \cite{Sznajd:2000,Galam:2002,Lorenz:2005}.

In most approaches to opinion dynamics, an agent's belief state is modelled as an opinion on either a single issue or multiple unrelated issues. We propose a model for opinion dynamics in which an agent's belief state consists of multiple interrelated beliefs. Because of this interconnectedness of the agent's beliefs, his belief state may be \emph{inconsistent}.
Consider the statement ``It is raining and it is not raining''. Even if one has no information on the current weather, one can see that this statement is false: its logical form is a contradiction, which is always false. Likewise, if one considers several aspects of the world simultaneously, one of the resulting theories about the world---to be made precise below---can be rejected out of hand as being logically inconsistent.

Our goal is to study the probability that an agent ends up with an inconsistent belief state. In order to explain how the beliefs are interrelated, how the agents revise or update their belief state, and how this leads to the possibility of inconsistency, we briefly review six aspects of our model: the content of an agent's opinion, the update rule according to which agents adjust their own opinion upon interaction with others, aspects related to the opinion profile, the time parameter, the group size, and the main research question. The notions introduced in the current section will receive a more formal treatment further on.

We compare our model with earlier approaches, in particular with the arguably best-known model for studying opinion dynamics, to wit, the Hegselmann--Krause (HK) model \cite{HegselmannKrause:2002}. In this model, the agents are trying to determine the value of an unspecified parameter and only hold one belief about this issue at any point in time. In the most basic version of the HK model, an agent updates his belief over time by averaging the beliefs of all those agents who are are within his `bound of confidence'; that is, the agents whose beliefs are not too distant from his own. The model we present can be regarded as an extension of the HK model.

\subsection{Content of an agent's opinion}\label{subsec:content}
Agent-based models of opinion dynamics come in two flavors: there are discrete and continuous models. In discrete models \cite{Sznajd:2000}, an agent's belief is expressed as a bit, 0 or 1 (\textit{cf.}\ Ising spin model in physics). The opinion may represent whether or not an agent believes a certain rumor, whether or not he is in favor of a specific proposal, whether or not he intends to buy a particular product, or which party the agent intends to vote for in a two-party system. This binary system can be generalized into discrete models that allow for more than two belief states (\textit{cf.}\ multi-spin
or Potts spin), which makes it possible to model multiple attitudes towards a single alternative or to represent preferences among multiple options \cite{Bernardes_etal:2001,Stauffer:2002a}. In continuous models \cite{Deffuant_etal:2000,HegselmannKrause:2002}, the agents each hold a belief expressed as a real number between 0 and 1. This may be used as a more fine-grained version of the discrete models: to represent the agent's attitude towards a proposal, a political party, or the like. In such models, values below 0.5 represent negative attitudes and values above 0.5 are positive attitudes. Alternatively, the continuous parameter may be employed to represent an agent's estimation of a current value or a future trend.

These models can be made more realistic by taking into account sociological and psychological considerations. For instance, they may be extended in a straightforward manner to describe agents who hold beliefs on multiple, independent topics (such as economic and personal issues \cite{Sznajd:2005}). As such, the models can account for the observation that agents who have similar views on one issue (for instance, taste in music) are more likely to talk about other matters as well (for instance, politics) and thus to influence each other's opinion on these unrelated matters \cite{Axelrod:1997}.\cite{fn:cont1} Nevertheless, it has been pointed out in the literature that these models are limited in a number of important respects, at least if they are to inform us about how real groups of agents interact with one another \cite{Douven:2010,DouvenRiegler:2010}. One unrealistic feature of the current models is that the agents only hold independent beliefs, whereas real agents normally have much richer belief states, containing not only numerous beliefs about possibly very different matters, but also beliefs that are logically interconnected.

In the discrete model that we propose, the belief states of the agents no longer consist of independent beliefs; they consist of \emph{theories} formulated in a propositional language (as will be explained in Section~\ref{subsec:LogicalFramework}). We will show that this extension comes at a cost. Given that the agents in earlier models hold only a single belief, or multiple, unrelated beliefs, their belief states are automatically self-consistent. This is not true for our model: some belief states consisting of interrelated beliefs are inconsistent.

\subsection{Update rule for opinions}\label{subsec:updaterule}
The update rule specifies how an agent revises his opinion from one point in time to the next. A popular approach is to introduce a \emph{bound of confidence}. This notion---which is also called `limited persuasion'---was developed first for continuous models, in particular the HK model \cite{HegselmannKrause:2002},\cite{fn:cf1} and was later applied to discrete models as well \cite{Stauffer:2002a}. Moreover, the idea of bounded confidence can be extended to update rules for belief states which are theories: such an HK-like update rule will be incorporated into our current model.

There is some empirical evidence for models involving bounded confidence. In a psychological experiment, Byrne \cite{Byrne:1961} found that when an agent interacts with another agent, the experience has a higher chance of being rewarding and thus oft leading to a positive relationship between the two when their attitudes are similar, as compared to when their attitudes differ. According to this `Similarity Attraction Paradigm', in future contacts, people tend to interact more with people who hold opinions similar to their own. Despite this evidence, some readers may not regard updating under bounded confidence as a natural way for individuals to adjust their opinions in spontaneous, face-to-face meetings. Those readers may regard the agents as experts who act as consultants in a Delphi study.\cite{fn:delphi} In
such a setting, the agents do not interact directly, but get feedback on each other's opinions only via a facilitator. When the facilitator informs each expert only of the opinion of those other experts that are within their bound of confidence, an HK-like update rule seems to apply naturally.

\subsection{Opinion profile}\label{subsec:opinionprofile}
An opinion profile is a way to keep track of how many agents hold which opinion. This can be done by keeping a list of names of the agents and writing each agent's current opinion behind his name. An anonymous opinion profile can be obtained by keeping a list of possible opinions and tallying how many agents currently hold a opinion; we will employ the latter type of profile. Opinion dynamics can be defined as the study of the temporal evolution of opinion profiles.

\subsection{Time}\label{subsec:time}
Many studies in opinion dynamics investigate the evolution of opinion profiles in the long run. Usually, a fixed point or equilibrium state is reached. Hegselmann and Krause, for instance, investigate whether iterated updating will ultimately lead groups of agents to full or partial consensus \cite{HegselmannKrause:2002}. Mas also investigates consensus- versus cluster-formation, as a function of the sociological make-up of the group under consideration \cite{Mas:2010}.

For sociologists, the behavior of opinion profiles at intermediate time steps may be more relevant than its asymptotic behavior. Research on voting behavior, for example, should focus on intermediate time steps \cite{Bernardes_etal:2002}; after all, elections take place at a set date, whether or not the opinion profile of the population has stabilized at that point in time.

In our study, we calculate the probability that an agent comes to hold an inconsistent opinion by updating. We do not investigate the mid- or long-term evolution of the opinion profile, but focus on the opinion profiles resulting from the very first update. In other words, we consider the opinion profile at only two points in time: the initial profile and the profile resulting from one round of updates.

\subsection{Group size}\label{subsec:groupsize}
Another interesting parameter to investigate in opinion dynamics is the group size. We are interested in updates which lead to inconsistent opinions, which may occur already for groups as small as three agents (see Section~\ref{subsec:DiscDil} below). The social brain hypothesis \cite{HillDunbar:2003} states that 150 relations is the maximum people can entertain on average: Lorenz presents this as an argument to model groups of agents of about this size \cite{Lorenz:2008}. Whereas this figure seems moderate from the sociological point of view, this is not necessarily the case from a mathematical viewpoint. As observed by Lorenz \cite{Lorenz:2008} (p.~323), ``[c]omplexity arises with finite but huge numbers of agents.'' Therefore, opinion dynamics is often studied in the limit of infinitely many agents, which makes it possible to express the equations in terms of `density of agents'. We will not do this in our current study: because of the previous observations, we should at least investigate the interval of 3 up to 150 agents.

\subsection{Research question}\label{subsec:researchquestion}
As we have remarked, the agents in our model may end up in an inconsistent belief state, even when all agents start out holding a consistent theory. The main question to be answered in this paper is: how \emph{likely} is it that this possibility will materialize? More exactly, we want to know what the probability is that an agent will update to an inconsistent belief state and how this probability depends on the number of atomic sentences in the agents' language and on the size of their community. To this end, an analytical expression is given and evaluated numerically, both exactly and using statistical sampling. It is shown that, in our model, an agent always has a probability of less than 2\,\% of ending up in an inconsistent belief state.
Moreover, this probability can be made arbitrarily small by increasing the number of atomic sentences or by increasing the size of the community.


\section{Preliminaries}\label{sec:prelims}
In this section, we first present the logical framework we assume throughout the paper. Then we specify the representation of the opinion profile and the employed update rule. Finally, we relate our work to previous research on judgment aggregation and the discursive dilemma.

\subsection{Logical framework}\label{subsec:LogicalFramework}

\subsubsection{Language and consequence relation}\label{subsubsec:language}
The agents in our model will have to judge a number of independent issues; we use the variable $M$ for this number (where $M\in\mathbb{N}^+$). Throughout this section, we will illustrate our definitions for the case in which $M=2$, the easiest non-trivial example. Each issue is represented by an atomic sentence. If the agents are bankers, the issues may be investment proposals; if they are scientists, the issues may be research hypotheses. As an example, one atomic sentence could be `Magnetic monopoles exist', and another `It will rain tomorrow'. Atomic sentences can be combined using three logical connectives: `and', `or', and `not'. The collection of sentences that can be composed in this way is called the language~$L$.

We assume a classical consequence relation for the language, which, following standard practice, we denote by the symbol $\vdash$. If $A$ is a subset of the language (a set of sentences) and $a$ is an element of the language (a particular sentence), then $A \vdash a$ expresses that $a$ is a logical consequence of $A$. That the consequence relation is classical means that it obeys the following three conditions: (1) if $a \in A$ then $A \vdash a$; (2) if $A \vdash a$ and $A \subseteq B$ then $B \vdash a$; and (3) if $A \vdash a$ and for all $b \in A$ it holds that $B \vdash b$, then $B \vdash a$. Semantically speaking, that $a$ is a logical consequence of $A$ means that, necessarily, if all the sentences in $A$ are true, then so is $a$.

\subsubsection{Possible worlds}\label{subsubsec:PossibleWorlds}
If we were to know which of the atomic sentences are true in the world and which are false, we would know exactly what the world is like (at least as far as is expressible in our language, which is restricted to a finite number of aspects of the world). The point is that our agents do not know what the world is like. Any possible combination of true--false assignments to all of the atomic sentences is a way the world may be, called a \emph{possible world}.

Formally, a possible world is an assignment of truth values to the atomic sentences. Hence, a language with $M$ atomic sentences allows us to distinguish between $w_{\max} = 2^M$ possible worlds: there is exactly one possible world in which all atomic sentences are true; there are $M$ possible worlds in which all but one of the atomic sentences are true; there are $\binom{M}{2}$ possible worlds in which all but two of the atomic sentences are true; and so on.

We may represent a possible world as a sequence of bits (bit-string). First we have to decide on an (arbitrary) order of the atomic sentences. In the bit-string, 1 indicates that the corresponding atomic sentence is true in that world, 0 that it is false. Let us illustrate this for the case in which there are $M=2$ atomic sentences: call `Magnetic monopoles exist' atomic sentence $m=0$ and `It will rain tomorrow' atomic sentence $m=1$. Then there are $w_{\max}=4$ possible worlds, $w \in \{0,\ldots,3\}$, which are listed in Table~\ref{table:ExampleWorlds}. Also the numbering of the possible worlds is arbitrary, but for convenience we read the sequence of 0's and 1's as a binary number. The interpretation of possible world
$w=2$, for example, is that sentence $m=0$ is false and sentence $m=1$ is true: in this possible world, it holds that magnetic monopoles do not exist and that it will rain tomorrow.

\begin{table}[!tb]
\caption{With $M=2$, there are $w_{\max}=2^M=4$ possible worlds, $w=0,\ldots,w=3$.}\label{table:ExampleWorlds} \centering \vspace{0.3cm}
\begin{ruledtabular}
\begin{tabular}{rcc}
 \hline\noalign{\smallskip}
  & \quad $m=1$ \quad & \quad $m=0$ \quad  \\
 \noalign{\smallskip}\hline\noalign{\smallskip}
  $w=0$ & 0 & 0 \\
  $w=1$ & 0 & 1 \\
  $w=2$ & 1 & 0 \\
  $w=3$ & 1 & 1 \\
 \noalign{\smallskip}\hline
\end{tabular}
\end{ruledtabular}
\end{table}

\subsubsection{Theories}\label{subsubsec:Theories}
A theory is a subset of possible worlds.\cite{fn:deftheorem} Let us explain this: an agent believes the actual world to be among the possible worlds that are in his theory; he has excluded the other possible worlds as live possibilities. To see that a theory may contain more than one specific possible world, consider an agent who is sure that `Magnetic monopoles exist' is false, but has no idea whether `It will rain tomorrow' is true or false. If these are the only atomic sentences in
his language, the agent holds a theory with two possible worlds. Given that we can order the possible worlds, we can represent theories as sequences of 0's and 1's, which in turn can be read as binary numbers. (This procedure is similar to the one used above for representing possible worlds by binary numbers.) Note that there are $t_{\max}=2^{w_{\max}}$ theories that can be formulated in a language with~$M$ atomic sentences.

Table~\ref{table:ExampleTheories} below illustrates this set-up for the case where $M=2$. In that table, theory $t=0$ is the inconsistent theory, according to which all worlds are impossible; syntactically, it corresponds to a contradiction. We know beforehand that this theory is false: by ruling out all possible worlds, it also rules out the actual world. Theory $t=15$ regards all worlds as possible; syntactically, it corresponds to a tautology. We know beforehand that this theory is true---the actual world must be among the ones that are possible according to this theory---but precisely for that reason the theory is entirely uninformative. The other theories are all consistent and of varying degrees of informational strength. The most informative ones are those according to which exactly one world is possible; a little less informative are those according to which two worlds are possible; and still less informative are the theories according to which three worlds are possible.

In Table~\ref{table:ExampleTheories}, we have numbered the theories by interpreting their bit-string notation as a binary number. The reverse order of the worlds in the top line is so as to make world $w$ correspond with the $w$\textsuperscript{th} bit of the binary representation of the theory.

\begin{table*}[!tb]
\centering \caption{With $M=2$, there are $w_{\max}=2^M=4$ possible worlds, $w=0,\ldots,w=3$, and $t_{\max}=2^{w_{\max}}=16$ different theories, $t=0,\ldots,t=15$. The penultimate column gives the sum of bits (bit-sum), $s_t$, of each theory. The last column represents the opinion profile of the community.}\label{table:ExampleTheories} \vspace{0.3cm}
\begin{tabular}{rcccccc}
 \hline\noalign{\smallskip}
   & \quad $w=3$ \quad & \quad $w=2$ \quad & \quad $w=1$ \quad & \quad $w=0$ \quad & \, $s_t$ \, & opinion profile \\
 \noalign{\smallskip}\hline\noalign{\smallskip}
  $t=0$ & 0 & 0 & 0 & 0 & 0& $n_0$\\
  $t=1$ & 0 & 0 & 0 & 1 & 1& $n_1$\\
  $t=2$ & 0 & 0 & 1 & 0 & 1& $n_2$\\
  $t=3$ & 0 & 0 & 1 & 1 & 2& $n_3$\\
  $t=4$ & 0 & 1 & 0 & 0 & 1& $n_4$\\
  $t=5$ & 0 & 1 & 0 & 1 & 2& $n_5$\\
  $t=6$ & 0 & 1 & 1 & 0 & 2& $n_6$\\
  $t=7$ & 0 & 1 & 1 & 1 & 3& $n_7$\\
  $t=8$ & 1 & 0 & 0 & 0 & 1& $n_8$\\
  $t=9$ & 1 & 0 & 0 & 1 & 2& $n_9$\\
  $t=10$ & 1 & 0 & 1 & 0 & 2& $n_{10}$\\
  $t=11$ & 1 & 0 & 1 & 1 & 3& $n_{11}$\\
  $t=12$ & 1 & 1 & 0 & 0 & 2& $n_{12}$\\
  $t=13$ & 1 & 1 & 0 & 1 & 3& $n_{13}$\\
  $t=14$ & 1 & 1 & 1 & 0 & 3& $n_{14}$\\
  $t=15$ & 1 & 1 & 1 & 1 & 4& $n_{15}$\\
 \noalign{\smallskip}\hline
\end{tabular}
\end{table*}

\subsection{Opinion profile}\label{subsec:OpinionProfile}
So far, we have focused on the belief state of a single agent, which is expressed as a theory. Now, we consider a community of~$N$ agents. The agents start out with (possibly different) information or preferences, and therefore may vote for different theories initially. The only assumption we make about the agents' initial belief states is that they are consistent. Subsequently, the agents are allowed to communicate and adjust there preference for a theory accordingly. In particular, we model what happens when the agents communicate with all other agents whose belief states are `close enough' to their own---that are within their bound of confidence, in Hegselmann and Krause's terminology---and update their belief state by
`averaging' over the close enough belief states, where the relevant notions of closeness and averaging are to receive formally precise definitions. The totality of belief states of a community at a given time can be represented by a string of $t_{\max}$ numbers, $n_0$, \dots, $n_{t_{\max}-1}$, where the number $n_t$ indicates how many agents hold theory $t$ at that time. We may also represent these numbers as a vector, $\overrightarrow{n}$. We refer to this string or vector as the (anonymous) opinion profile of the community at a specified time. Because each agent has exactly one belief state, the sum of the numbers in an opinion profile is equal to the total number of agents, $N$. Also, given that initially no agent has the inconsistent theory as his belief state, $n_0$ is always zero before any updating has taken place. Later this may change. By updating, an agent may arrive at the inconsistent theory; we shall call such an update a \emph{zero-update} (because the inconsistent theory is represented by a string of only 0's).

In most opinion dynamics studies, a random opinion profile is used as a starting point. Because our question deals with a probability in function of the initial opinion profile, we explicitly take into account \emph{all possible} initial opinion profiles, or---where this is not possible---take a large enough statistical sample out of all possible initial opinion profiles. The different opinion profiles can be thought of as resulting from the individual choices the agents make regarding which world or worlds they deem possible. Here, we assume that the adoption of a theory as an initial belief state can be modeled as a sequence of $2^M$ independent tosses of a fair coin, where the agent is to repeat the series of tosses if the
result is a sequence of only 0's. As a consequence, all consistent theories have the same probability---namely, $1/(t_{\max}-1)$---of being adopted as an initial belief state by an agent. That is to say, we are studying what in the literature are sometimes referred to as `impartial cultures' (\textit{cf.}~Section~\ref{subsec:DiscDil}). Furthermore, the agents are assumed to choose independently of each other.

\subsection{Update rule}\label{subsec:UpdateRule}
Theorists have studied a variety of update rules, depending on the application the authors have in mind. For instance, to model gossip communication, Deffuant \textit{et al.}~use a rule in which updates are triggered by pairwise interactions \cite{Deffuant_etal:2000}. To model group meetings, the updates should rather be simultaneous within the entire group of agents. During a conference, the agents meet each other face-to-face; in that case, additional effects should be taken into account, such as the `primacy effect', which demonstrates that the order in which the agents' opinions are publicly announced may influence how the others revise their opinion.

As mentioned before, we may think of our group of agents as a group of scientists, bankers, or other experts who act as consultants in a Delphi-study. The choices in the selection of the update rule follow from that. Delphi-studies are typically conducted in a way such that the experts do not have any direct interaction \cite{LinstoneTuroff:1975}. Thus, we need a model with simultaneous updating but without primacy effects: in this respect, the update rule of the HK model \cite{HegselmannKrause:2002} applies to this situation in a natural way.

Another relevant aspect of the HK model is that an agent may not take into account the opinions of \emph{all} the agents in the group. This may occur when the agent knows all the opinions but does not want to take into account the opinions of agents who hold a view that is too different from the agent's own, or because the facilitator of the Delphi-study only informs the agent about the opinions of experts who hold an opinion similar to the agent's.

In order to quantify what counts as a similar opinion, we introduce the `maximal distance' or `bound of confidence', $D$. This parameter expresses the number of bits that another agent's opinion may maximally differ from one's own if that agent's opinion is to be taken into account in the updating process. To quantify the difference between two theories, we use the so-called Hamming distance of the corresponding bit-strings, defined as the number of digits in which these strings differ \cite{Hamming:1950}.

It is possible to consider heterogeneous populations, where agents may have different bounds of confidence \cite{HegselmannKrause:2005}. Because Hegselmann and Krause report no qualitative difference between the homogeneous and the heterogeneous case \cite{HegselmannKrause:2005}, we choose the simpler, homogeneous approach: $D$ has the same value for all agents in any population we consider. We investigate the influence of the value of $D$ on the probability of updating to the inconsistent theory. By an agent's `neighbors' we refer to the agents whose opinions fall within the bound of confidence of the given agent. Note that, however $D$ is specified, an agent always counts as his or her own neighbor.

At this point, we still have to specify \emph{how} agents update on the basis of their neighbors' belief states. Like Hegselmann and Krause in most of their studies \cite{HegselmannKrause:2002,HegselmannKrause:2006,HegselmannKrause:2009}, we choose the arguably simplest and also the most plausible averaging method, which is to determine an agent's new belief state based on the straight average of his neighbors' belief states. Our update rule for theories is a bitwise operation in two steps---averaging and rounding. First, each bit is averaged by taking into account the value of the corresponding bit of an agent's neighbors. In general, the result is a value in $[0,1]$ rather than in $\{0,1\}$. Hence the need for a second step: in case the average-of-bits is greater than $\frac{1}{2}$, the corresponding bit is updated to 1; in case the average-of-bits is less than $\frac{1}{2}$, the corresponding bit is updated to 0; and in case the average-of-bits is exactly equal to $\frac{1}{2}$, the corresponding bit keeps its initial value.\cite{fn:defudm}

In the current study, we are only interested in the probability of arriving at an inconsistent conclusion after a single update. However, again following Hegselmann and Krause, one could also designate one of the theories expressible in the agents' language as the truth and allow the agents to gather evidence which points toward the truth. One could then study the interplay between convergence to the truth and avoiding inconsistencies. We plan to implement this in future research.

We want to keep the model as simple as possible. Therefore, we will not implement any of the following additional parameters: trustworthiness of agents, physical closeness and/or social network \cite{Galam:2002}, and other psychologically relevant aspects (such as bias, self-justification, and getting tired of repeated updating \cite{TavrisAronson:2007}). Because we only have one update rule, there is no need to consider mixed groups and/or agents changing their update rule over time, which would complicate matters even further. In general, pure cases have the drawback of being less realistic, at the benefit of showing more clearly the effect of a single parameter.

\subsection{Comparison with related work}\label{subsec:DiscDil}
The possibility of inconsistent outcomes resulting from a voting procedure, similar to updating beliefs, has already received some discussion in the literature on the so-called discursive dilemma. According to majority voting on a set of interrelated propositions, a proposition should be made part of the collective judgment if it is part of a majority of the individual judgments. The discursive dilemma \cite{Pettit:2001,ListPuppe:2007}, which is a more general form of the doctrinal paradox \cite{KornhauserSager:1986}, shows that this voting procedure may result in an inconsistent collective judgment even if all the individual judgments are consistent. This is immediately relevant to our present concerns, given that majority voting falls under the definition of averaging to be employed in our update rule.

The original example of the doctrinal paradox is stated in the context of a legal court decision. It presents three judges who have to vote on three propositions: two premisses $P$~and~$Q$, and a conclusion~$R$. The connection between the premisses and the conclusion is motivated by legal doctrine and formalized as $R \leftrightarrow (P \wedge Q)$. A vote is called consistent if it satisfies this rule, and inconsistent otherwise. Readers will have no difficulty assigning consistent individual judgments to the judges which, given proposition-wise majority voting, nevertheless give rise to an inconsistent collective judgment.

In \cite{Pettit:2001}, it is argued that this type of paradoxical result can occur in group decisions on interrelated propositions in other contexts as well. The example can be generalized to cases with more than two premises \cite{Pettit:2001}, more than three judges \cite{List:2005}, or to cases with a disjunctive connection rule $R \leftrightarrow (P \vee Q)$ rather than a conjunctive one \cite{List:2005}. There are many impossibility results to be found in the literature \cite{ListPuppe:2007}, that show that---given some plausible conditions---there is no way of aggregating consistent individual judgments so as to guarantee a consistent group judgment.

List \cite{List:2005} (p.~5) addresses the question of how serious the threat posed by this paradox is. He gives a probabilistic analysis of the discursive dilemma. In the general case of $k$ premisses, $P_1,\ldots,P_k$, and one conclusion, $R$, there are $2^k$ combinations of the premisses being true or false (possible worlds). Each time, the truth or falsehood of the remaining proposition, the conclusion, is determined by the conjunctive connection rule $R \leftrightarrow (P_1 \wedge \cdots \wedge P_k)$. The simplest case is that in which all the probabilities are equal that an agent holds a particular opinion out of the $2^k$ consistent ones; this situation is called an \emph{impartial culture}. List \cite{List:2005} considers this case as well as variants thereof. In his paper, he also analyzes the case of an \emph{impartial anonymous culture}, which
takes every anonymous opinion profile to be equally likely (rather than every individual choice of the agents). Following the literature on the Condorcet jury theorem, he assumes identical probabilities for all agents and independence between different agents. The focus of \cite{List:2005} is mainly on convergence results (in particular, on the probability of inconsistency in the limit of the number of agents going to infinity), although some results are stated in terms of a finite (but always odd) number of agents.

While we also intend to give a probabilistic analysis of the occurrence of inconsistencies in what may be interpreted as group judgments and make the result general for the number of atomic propositions considered by the agents, there are some differences between our model and that of List \cite{List:2005} that merit highlighting.

First, we want to model agents that---in the terminology of the discursive di\-lem\-ma---vote on a theory. The type of inconsistencies encountered in the doctrinal paradox can be avoided relatively easily by having the jurors vote either on the premisses only (and then derive the conclusion from the collective judgments on the premises) or on the conclusion only \cite{Pettit:2001}. For voting on theories, which are by definition closed under derivability, there is no quick fix available to avoid the problem that agents reach the inconsistent theory by majority voting (or, in our terms, updating by averaging). Therefore, the question regarding the probability of this event seems all the more pressing.

Second, we are interested in calculating the probability of an inconsistency for a completely general number of agents (odd as well as even), rather than in convergence results (in the limit of infinitely many agents).

And third, rather than considering majority voting where the relevant majority has to be relative to the whole group of agents, we assume an update rule that admits of greater and smaller bounds of confidence, which effectively comes to requiring a majority only relative to a subgroup of agents. This also implies that if one agent comes to hold an inconsistent theory, this need not be so for all agents in the group.

On a more practical level, we note that, because we already take into consideration three parameters (number of atomic sentences, number of agents, and the bound of confidence parameter related to the update rule), we confine our discussion to impartial cultures.

\section{The probability of inconsistencies}\label{sec:Analytical}
We now turn to the question of how probable it is that an agent with a consistent initial belief state updates to the inconsistent belief state by averaging the (also initially consistent) belief states of that agent's neighbors. More precisely, we consider a fixed update rule---a fixed way of averaging belief states---and study the effects on the said probability of the following parameters:

\medskip

\begin{compactenum}
   \item the number~$M$ of atomic sentences of the agents' language;
   \item the number~$N$ of agents in the community;
   \item the bound of confidence, $D$, which is the maximal Hamming distance for one agent to count as a neighbor of another.
\end{compactenum}

\medskip

\noindent The analytical solution consists of many nested sums. In the next section, we evaluate the analytical expression numerically. Because exact calculations are only feasible for small populations, we extend the calculations by a simulation based on statistical sampling.

Given $M$ atomic sentences, $N$ agents, and a maximal Hamming distance or bound of confidence $D$, we want to calculate, first, the fraction of agents who update to the contradiction, when we consider all agents in all possible initial opinion profiles, $F_{\textrm{AG}}(M,N,D)$; and second, the fraction of all possible opinion profiles that have at least one agent who updates to the contradiction, $F_{\textrm{OP}}(M,N,D)$. In other words, $F_{\textrm{AG}}(M,N,D)$ is the probability for an agent to update to the inconsistent theory in a single update under the assumption that nothing is known about the opinion profile---only the parameters $M$, $N$, and $D$ are known. Likewise, $F_{\textrm{OP}}(M,N,D)$ is the probability that at least one agent in the entire population will update to the inconsistent theory in a single update. Clearly, the latter probability should be at least as great as the former.

Readers who are interested in the details of the derivation of the analytical expressions for these probabilities are referred to the appendix. Here, we state the result of the derivation, introduce previously undefined parameters occurring in it, and clarify the overall form of the expressions for $F_{\textrm{AG}}(M,N,D)$ and $F_{\textrm{OP}}(M,N,D)$:

\begin{subequations}\label{eq:FractionsShorter}
\begin{multline}\label{eq:FractionsShorterAG}
F_{\textrm{AG}}(M,N,D)\:\: =
\\
\sum_{n_{1}=0}^N \sum_{n_{2}=0}^{N-n_{1}} \cdots
\sum_{n_{t_{\max}-2}=0}^{N-(n_{1}+n_{2}+\cdots+n_{t_{\max}-3})} \frac{N!}{n_{0}!n_{1}! \cdots
n_{t_{\max}-1}!} \:\: \times
\\
\frac{1}{(t_{\max}-1)^N} \sum_{t=0}^{t_{\max}-1} \left( \frac{n_{t}}{N} \prod_{w=0}^{w_{\max}-1}
\textrm{INV}\big[ \big\langle B_w(t) \big\rangle \big] \right);
\end{multline}
\begin{multline}\label{eq:FractionsShorterOP}
F_{\textrm{OP}}(M,N,D) \:\: =
\\
\sum_{n_{1}=0}^N \sum_{n_{2}=0}^{N-n_{1}} \cdots
\sum_{n_{t_{\max}-2}=0}^{N-(n_{1}+n_{2}+\cdots+n_{t_{\max}-3})} \frac{N!}{n_{0}!n_{1}! \cdots
n_{t_{\max}-1}!} \:\: \times
\\
\frac{1}{(t_{\max}-1)^N} \textrm{ZUP}(M,N,D,\overrightarrow{n}).
\end{multline}
\end{subequations}

\noindent Because these expressions take the form of nested sums, in order to explain them, we should start by looking at the last part, which is the actual core of the equation. The expression for the population-based fraction, $F_{\textrm{OP}}(M,N,D)$, is very similar to that for $F_{\textrm{AG}}(M,N,D)$ except for that most central part. First we look at the expression for the agent-based fraction, $F_{\textrm{AG}}(M,N,D)$.

At the heart of the expression for $F_{\textrm{AG}}(M,N,D)$, we find the function $\big\langle B_w(t) \big\rangle$, which specifies how a given agent in a fixed opinion profile updates the bits of his theory: it calculates the average of the $w$\textsuperscript{th} bit for an agent in opinion profile $\overrightarrow{n}$ with bound of confidence $D$ whose initial opinion is theory~$t$. Because we can do this for all bits, we can determine whether or not this agent updates to the inconsistent theory; the expression
\begin{equation}\label{eq:Central-F_AG}
\prod_{w=0}^{w_{\max}-1} \textrm{INV}\big[ \big\langle B_w(t) \big\rangle \big]
\end{equation}
\noindent evaluates to 1 if this is the case, and to 0 otherwise.

As a next step, we need to count the zero-updates for all the agents in the opinion profile, not just for one: $\sum_{t=0}^{t_{\max}-1}$ sums over all possible initial opinions and the factor $n_{t}$ takes into account how many agents hold each of these opinions initially. We divide by $N$ for normalization.

Moreover, we need to take into account all different, anonymous opinion profiles $\overrightarrow{n}$, not just a particular one:
\begin{equation*}
\sum_{n_{1}=0}^N \sum_{n_{2}=0}^{N-n_{1}} \cdots
\sum_{n_{t_{\max}-2}=0}^{N-(n_{1}+n_{2}+\cdots+n_{t_{\max}-3})}
\end{equation*}
\noindent sums over all possible anonymous opinion profiles. Thus, at the right-hand side of these summations, all the $n_t$'s have a fixed value, meaning that there, the full opinion profile, $\overrightarrow{n}$, is specified. The weight factor $\frac{N!}{n_{0}!n_{1}! \cdots n_{t_{\max}-1}!}$ takes into account that certain individual choices of agents result in the same anonymous opinion profile. With this weight factor, we consider an impartial culture; omitting it would result in an impartial \emph{anonymous} culture (see also \cite{List:2005}). The remaining factor $\frac{1}{(t_{\max}-1)^N}$ is yet another normalization factor: it divides the result by the number of different (non-anonymous) opinion profiles.

We have seen that for the agent-fractions, the central expression (\ref{eq:Central-F_AG}) calculates the fraction of agents in the particular opinion profile $\overrightarrow{n}$ which perform a zero-update. Let us now look at the opinion-based fractions: there, the central expression is replaced by
\begin{equation*}
\textrm{ZUP}(M,N,D,\overrightarrow{n}),
\end{equation*}
\noindent which evaluates to 1 if the corresponding agent-based term (\ref{eq:Central-F_AG}) is non-zero, and to 0 otherwise. Because the rest of the expression for $F_{\textrm{AG}}(M,N,D)$ and $F_{\textrm{OP}}(M,N,D)$ is identical, this ensures that $F_{\textrm{AG}}(M,N,D) \leq F_{\textrm{OP}}(M,N,D)$.

\section{Numerical evaluation of the probability of inconsistency}\label{sec:Numerical}

It is far from trivial to estimate the outcome of the expressions for $F_{\textrm{AG}}(M,N,D)$ and $F_{\textrm{OP}}(M,N,D)$ or to analyze their limiting behavior as $N$ and/or $M$ become large. To obtain an idea of the quantitative output and qualitative behavior of the formulae, we evaluate them numerically.

\subsection{Exact calculations}\label{subsec:Exact}
Because the number of computations required to evaluate Equation~(\ref{eq:FractionsShorter}) is considerable, we have written a computer program (in Object Pascal) capable of evaluating the expression for the number of atomic sentences $M=1$ to $M=3$.\cite{fn:largeM} Here we summarize the results.

\begin{figure*}[!tb]
\includegraphics[width=15cm,keepaspectratio=true]{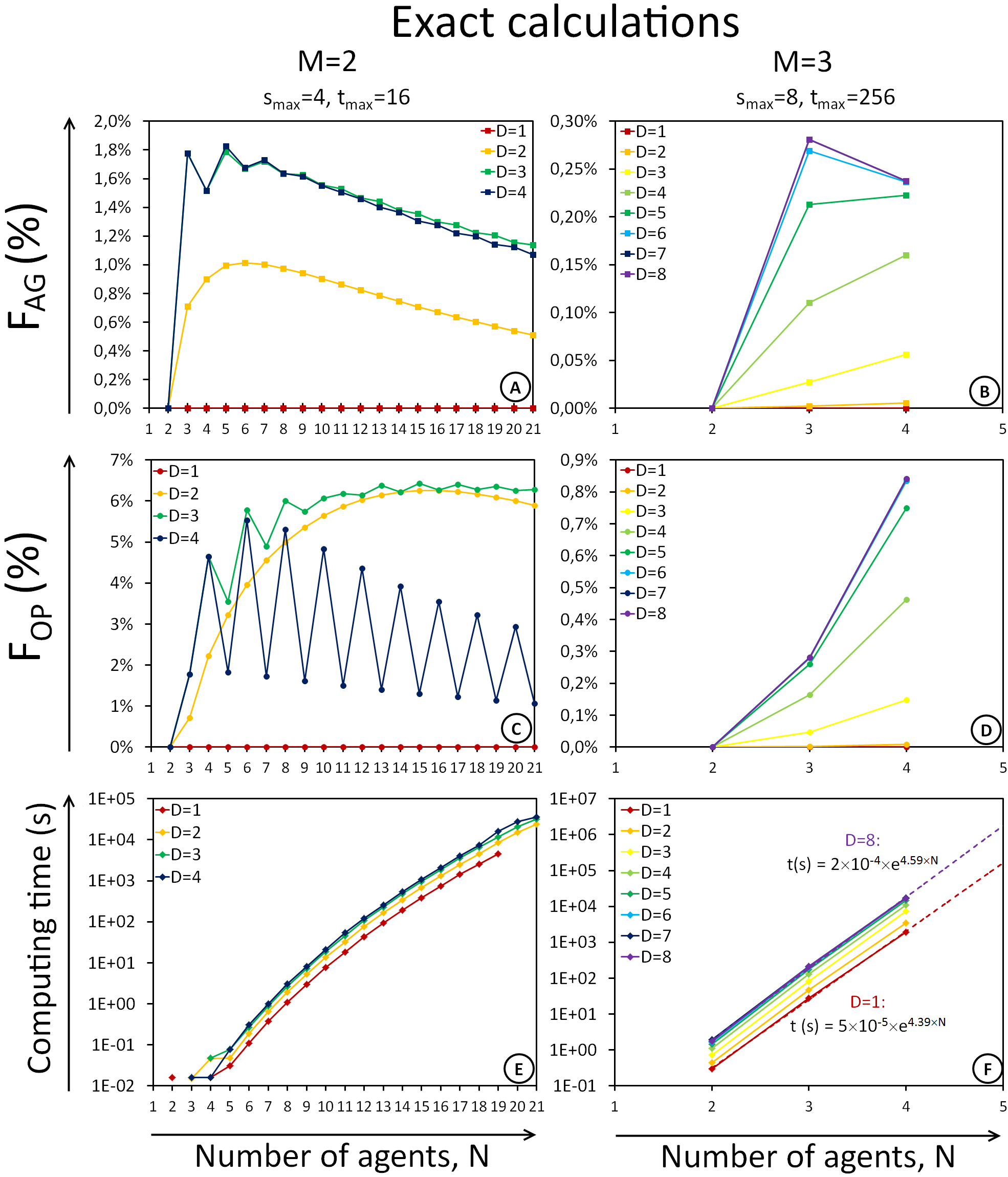}\\
  \caption{(Color online) Results of exact calculations. The number $M$ of atomic sentences equals 2 in the graphs at the left hand side (A, C, and E), and 3 in those at the right (B, D, and F). All graphs are presented as a function of the number of agents, $N$. The different curves in each graph represent different bounds of confidence, $D$. (A, B) Fraction (\%) of agents who update to the inconsistent theory. (C, D) Fraction (\%) of opinion profiles with at least one agent who updates to the inconsistent theory. (E, F) Computation time on one CPU in seconds on a semi-logarithmic scale.}\label{Fig:M2M3Exact}
\end{figure*}

If $M$ is equal to $1$, there are no opinion profiles in which any agent updates to the contradiction, no matter which values $N$~and~$D$ have. For $M=2$, we obtained exact results for $N=2$ up to $N=21$, as shown in Panels A and C of Figure~\ref{Fig:M2M3Exact}. For $M=3$, we could obtain exact results for $N=1$ up to $N=4$, as shown in Panels B and D of Figure~\ref{Fig:M2M3Exact}.

If $D=1$ or $N=2$, a zero-update never occurs. For $D>w_{\max}=2^M$, the number of zero-updates is equal to that for $D=2^M$ (data not shown). Because for values of $D = 2^M$ onward, all agents have all other agents as their neighbors, increasing $D$ further makes no difference.

The lowest values for $M$, $N$, and $D$ that may result in a zero-update are: $M=2$, $N=3$, and $D=2$. (In agreement with the doctrinal paradox, we find that a zero-update can occur for three agents.) This is a case that can be checked on paper, and can thus be used for testing the validity of the analytical expression and the program. Checking an example by hand is also a good way to get to understand the model better.

For $M=2$, we can consult Table~\ref{table:ExampleTheories}. We see that there are four theories with a bit-sum $s_t = 1$: $t=1$, $t=2$, $t=4$ and $t=8$. The distance, $d$, between any two of these theories equals~2. Consider an opinion profile with one agent holding $t=1$, one agent holding $t=2$, and one agent holding $t=4$. If $D\geqslant 2$, then all agents are each others' neighbors. Therefore, to update, they all take the same average: $\frac{1}{3} \big( (t=1)+(t=2)+(t=4) \big) =\bigl(0,\frac{1}{3},\frac{1}{3},\frac{1}{3}\bigr)$. After rounding, their new opinion becomes $(0,0,0,0)$; that is, they all arrive at $t=0$, the inconsistent theory. For $N=3$, populating three out of four of the theories $t=1$, $t=2$, $t=4$, and $t=8$ may happen in four different ways (given anonymity). If we keep track of the identity of the agents within each of these four opinion profiles, the agents may choose their belief states in six different ways, giving rise to $4\times 6=24$ non-anonymous configurations that lead at least one agent, and thereby in fact all agents, to the inconsistent theory in just one update. For $M=2$ and $N=3$, there are $3~375$ possible non-anonymous opinion profiles; with $D=2$, the aforementioned 24 opinion profiles are the only starting points from which to arrive at the inconsistent theory. Therefore, the opinion-profile-fraction is $F_{\textrm{OP}}(2,3,2) = 24/3~375 = 0.711\,\%$. Because all three agents update to the inconsistent theory, the agent-fraction is exactly equal to this: $F_{\textrm{AG}}(2,3,2) = 0.711\,\%$. These results are identical to the calculated value represented in the graphs.

For $D=3$ and $D=4$, the previous four anonymous (or 24 non-anonymous) configurations still lead to a zero-update, but there are additional possibilities which lead to the same result, to wit, those in which one agent occupies one of the six theories that have bit-sum $s_t = 2$ (namely, $t=3$, $t=5$, $t=6$, $t=9$, $t=10$, and $t=12$) and the other two agents each occupy a theory of bit-sum~1 such that the single 1-bit of the latter corresponds with a 0 in the first agent's theory. For instance, the combination of one agent holding $t=3$ with another holding $t=4$ and the third holding $t=8$ leads to an average of $\frac{1}{3}\big((t=3)+(t=4)+(t=8)\big)=\bigl(\frac{1}{3},\frac{1}{3},\frac{1}{3},\frac{1}{3}\bigr)$, which becomes $t=0$ after rounding. As is readily seen, higher bounds of confidence and larger population sizes soon become too complex to check by hand. That is why computer calculations are indispensable for this type of research.

As for agent fractions, for $M=2$ (Panel A of Figure~\ref{Fig:M2M3Exact}), all curves have a similar peak shape. The values for $D=3$ and $D=4$ are very similar, and almost double as compared to those for the smaller bound of confidence $D=2$. The former curves exhibit an `odd--even wobble' near the top: for an odd number of agents a larger fraction of the population updates to the inconsistent theory than one would expect based on the behavior of even-numbered groups of similar size.

For $M=3$ (Panel B), it is clear again that higher maximal distances give rise to higher fractions. Also the odd--even wobble seems present, especially for higher maximal distances, but we need more data to confirm this.

As for opinion profile fractions, for $M=2$ (Panel C), the curves $D=3$ and $D=4$ again exhibit an odd--even wobble. Curiously, here the trend is opposite to that observed in Panel B: for an even number of agents a larger fraction of the population updates to the inconsistent theory than one would expect based on the behavior of odd-numbered groups of similar size. Furthermore, $D=3$ and $D=4$ are not as similar as is the case in panel B: for $D=3$, the wobble decreases as the curve attains a maximum or plateau, whereas for $D=4$ the wobble only decreases along with the overall amplitude decrease of the curve.

The onset of the curves for $M=3$ (Panel D) seems to indicate a similar reversed odd--even wobble, but again we need more data to confirm this.

Finally, a word on computing time: Although exact results for higher numbers of agents than shown here are attainable in principle, they come with ever increasing computational costs. Panels E and F of Figure~\ref{Fig:M2M3Exact}---note the log-scale on the vertical axis---show that the required computing time increases nearly exponentially with the number of agents. For $M=3$, we have extrapolated the computing time to $N=5$ for $D=1$ and $D=8$, and found that one additional data point would require about 5 (for $D=1$) to 21 (for $D=8$) days of computation. Of course, these results are machine-dependent, but the exponential trend is intrinsic, since it is related to the fast increase in the number of terms in Equation~(\ref{eq:FractionsShorter}). Obtaining more data is thus limited by practical constraints, unless we approach the problem differently.

\subsection{Extending the numerical analysis by statistical sampling}\label{subsec:Statistical}

Instead of constructing all possible opinion profiles and counting how many agents update to the inconsistent theory, we now consider a statistical approach: we have a\-dap\-ted the program used for calculating the exact results of the previous section to draw a random sample from all possible opinion profiles and calculate the fractions of agents and opinion profiles within the sample that update to the inconsistent theory. If the sample size is sufficiently large, these sample fractions are good estimates of the respective fractions in the complete set of opinion profiles.

\begin{figure*}[!tb]
\includegraphics[width=15cm,keepaspectratio=true]{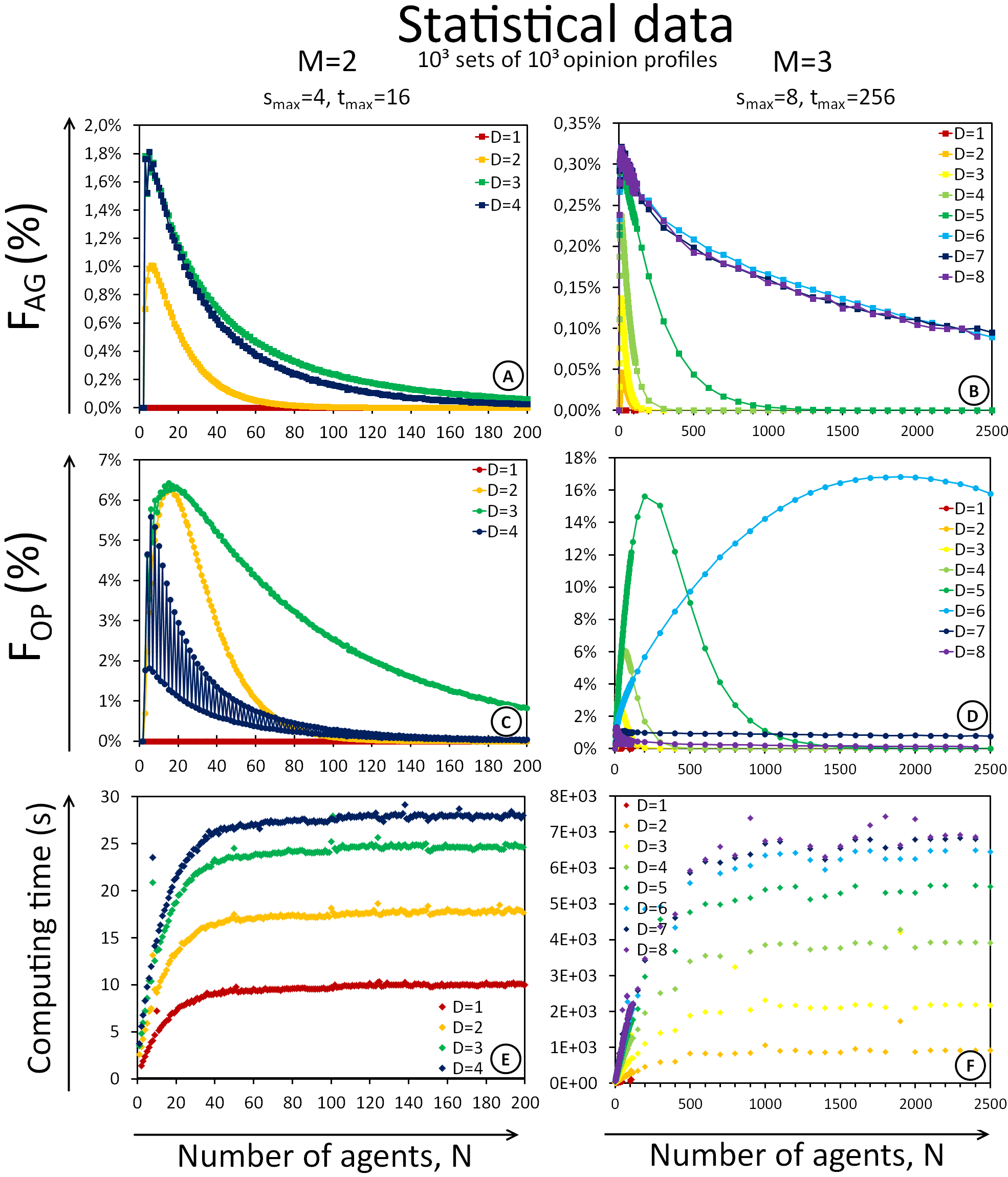}\\
  \caption{(Color online) Results of calculations based on statistical sampling, consisting of $1~000$ sets of $1~000$ opinion profiles each. (A, B) Fraction (\%) of agents who update to the inconsistent theory. (C, D) Fraction (\%) of opinion profiles with at least  one agent who updates to the inconsistent theory. (E, F) Computation time on one CPU in seconds (linear scale). The number $M$ of atomic sentences is two in the graphs at the left hand side (A, C, and E), and three in those at the right (B, D, and F). All graphs are presented as a function of the number~$N$ of agents. The different curves in each graph represent different bounds of
  confidence, $D$.}\label{Fig:M2M3Stat}
\end{figure*}

We have done tests with different sample sizes, looking for a good trade-off between low noise on the data and acceptably low computational costs. All results presented in Figure~\ref{Fig:M2M3Stat} were obtained using samples of $10^3$ sets of $10^3$ opinion profiles each, that is $10^6$ opinion profiles in total per data point. (Smaller sample sizes such as $10^4$ opinion profiles per data point require a computation that is 100 times faster, but produce curves that are visibly noisy.)

Because we have some exact results, we can use these to assess the statistical program: the onset of the curves in panels A--D of Figure~\ref{Fig:M2M3Stat} corresponds well with the data presented in the respective panels of Figure~\ref{Fig:M2M3Exact}.

As can be seen in panels E and F of Figure~\ref{Fig:M2M3Stat}, the computation time increases at first, but then remains almost constant: a typical calculation no longer depends on the number of agents, $N$, but only on $M$ and $D$. (The outliers are due to periods of standby time of the computer that was used for the calculations.) Thus, the approach with statistical sampling makes it possible to investigate the curves up to a much higher value of $N$ than using the exact formula.

We have plotted the curves for $M=2$ up to $N=200$, where all the curves have long passed their maximum and are decreasing smoothly. For $M=3$, we have plotted the curves up to $N=2~500$, because the maximum for $D=6$ is only obtained at around $N=1~800$. Because the curves are smooth (apart from the odd--even wobble), instead of computing every point, from $N=110$ onwards we have increased the step size to $\Delta N=100$.

First let us first examine the odd--even wobble that we noticed in the onset of the curves from the exact calculations.

For $M=2$, in Figure~\ref{Fig:M2M3Stat}.A we see that for the agent-based fractions, the oscillation is only present for low $N$-values. In Figure~\ref{Fig:M2M3Stat}.C, we see that for $D=2$ there is no odd--even wobble. For $D=3$, it is present before the maximum in the curve, but not beyond it. For $D=4$, the oscillation is well pronounced throughout the curve, although the amplitude of the oscillation diminishes as the curve drops.

\begin{figure}[!tb]
\centering
  \includegraphics[width=8.0cm,keepaspectratio=true]{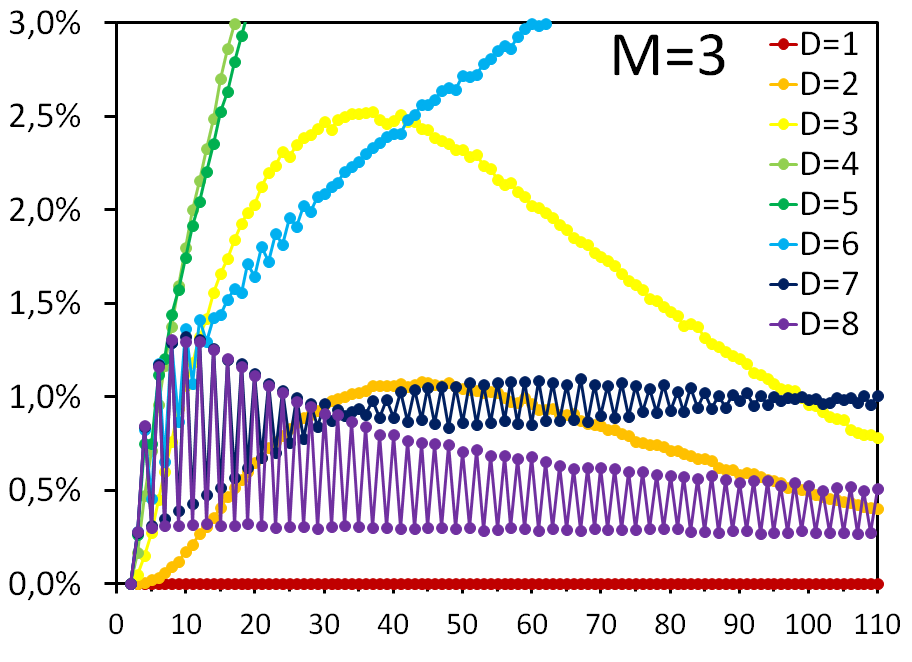}\\
  \caption{(Color online) Detail of Panel D in Figure~\ref{Fig:M2M3Stat}: fraction (\%) of opinion profiles with at least one agent who updates to the inconsistent theory in the case where $M=3$. }\label{Fig:M3StatDetail}
\end{figure}

\begin{figure}[!tb]
\centering
  \includegraphics[width=8.0cm,keepaspectratio=true]{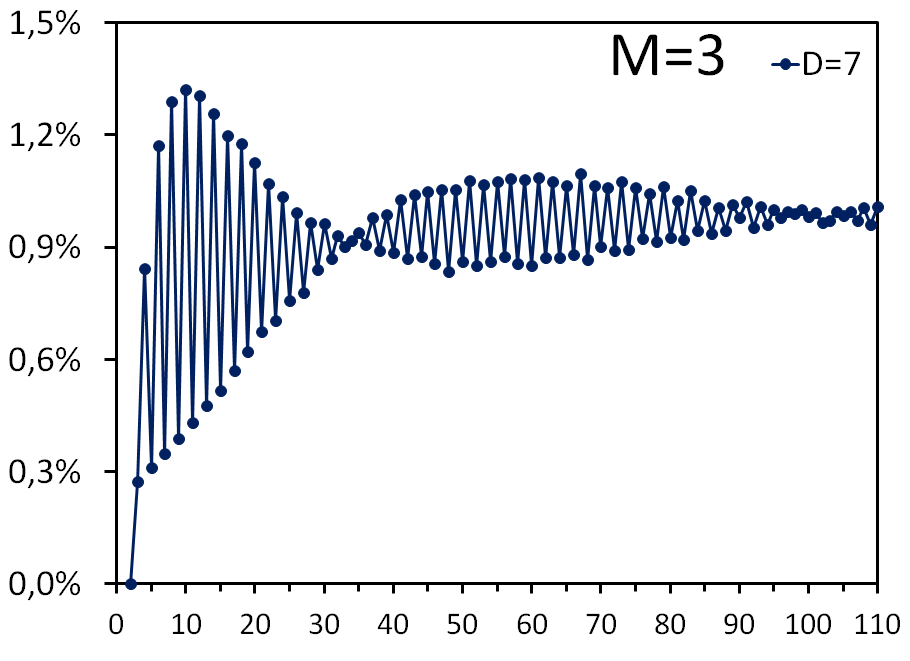}\\
  \caption{(Color online) Separate graph of maximal distance $D=7$: fraction (\%) of opinion profiles with at least one agent who updates to the inconsistent theory in the case where $M=3$. }\label{Fig:M3StatDetailD7}
\end{figure}

For $M=3$, in the agent-based fractions in Figure~\ref{Fig:M2M3Stat}.B we see no odd--even wobble in the decreasing tails of the curves. Figure~\ref{Fig:M3StatDetail} provides a detail of the curves in panel~D for the region near the origin where all curves overlap. There is no wobble visible for $D \in \{2,\ldots,5\}$. For $D=6$, there is an oscillation for $N$ up to about $30$ (long before the curve attains its maximum). For $D=8$, the oscillation seems to go on for all values of $N$ (much like curve $D=4$ for $M=2$). Because the start of curve $D=7$ overlaps with that of $D=8$, we present the former in a separate graph (\textit{cf.}~Figure~\ref{Fig:M3StatDetailD7}). $D=7$ is the only case in which we can see an amplitude modulation (like that in interfering sound waves, where the phenomenon is known as `beats'): the oscillation seems to disappear at about $N=34$ but its amplitude increases again for larger $N$ until $N=100$. There, a new oscillation starts, but since we have lowered the sampling from thereon to $\Delta N=100$, we cannot examine it further.
We realize that this odd--even wobble cries out for an explanation. Currently, however, we have no conclusive explanation for it, so we present it as a puzzle.

Let us now consider the positions for which the various curves become maximal. This is a worst-case analysis, because the maximal fractions correspond to situations with the highest probability for an agent to obtain the inconsistent theory by following the studied update rule, and for a population to have at least one agent who updates to the inconsistent theory. In Table~\ref{table:max_agents} and Table~\ref{table:max_profiles}, the maximal fractions for $M=2$ and $M=3$ are listed in terms of agents and opinion profiles, respectively.\cite{fn:M2max} From the table, we see that compared
to $M=2$, for $M=3$ the maximum occurs for larger $N$. Moreover, the percentage at the maximum is much larger for the opinion-profile-based fractions, but much smaller for the agent-based ones.

At first blush, this may seem a strange combination, so let us explain what is going on here: looking at the set of all possible opinion profiles (for the same $N$), there is a certain number of them that contains at least one agent who updates to the inconsistent theory. However, as the population grows larger, the average number of agents who update to the inconsistent theory decreases. Zero-updates require `asymmetrical' opinion profiles, with all agents more or less evenly distributed at the lower bit-sum theories. (Recall the example for $M=2$ and $N=3$.) However, if there are many more agents than theories, $N \gg t_{\max}$, then there are many more combinations of individual choices that lead to a situation with a similar number of agents at each theory (low and high bit-sum) than there are combinations which produce opinion profiles with the agents primarily present at low-bit-sum theories. This is a consequence of the statistical law of large numbers: due to symmetry reasons, in the `average opinion profile' (each possible belief state instantiated by $N/(t_{\max}-1)$ agents) no zero-updates are possible and, as $N$ increases, the probability of an initial opinion profile being close to the average opinion profile also increases. Thus, large interacting groups act as a protective environment to keep the belief states of the group members consistent.

\begin{table}[!tb]
\caption{Maximal probability of obtaining the inconsistent theory after one update expressed as fraction (\%)
of agents.}\label{table:max_agents} \centering \vspace{0.3cm}
\begin{tabular}{cccl}
 \hline\noalign{\smallskip}
  ${}\:M\:$ & ${}\:D\:$ & ${}\:N\:$ & ${}\:F_{\textrm{AG}}$ (\%)\\
 \noalign{\smallskip}\hline\noalign{\smallskip}
  2 & 1 & / & /\\
  2 & 2 & 6 & 1.0081\\
  2 & 3 & 3 & 1.7824\\
  2 & 4 & 5 & 1.8112\\
 \noalign{\smallskip}\hline\noalign{\smallskip}
  3 & 1 & / & / \\
  3 & 2 & 22 & 0.0462\\
  3 & 3 & 14 & 0.1368\\
  3 & 4 & 11 & 0.2380\\
  3 & 5 & 15 & 0.2986\\
  3 & 6 & 13 & 0.3210\\
  3 & 7 & 19 & 0.3215\\
  3 & 8 & 13 & 0.3202\\
 \noalign{\smallskip}\hline
\end{tabular}
\end{table}

\begin{table}[!tb]
\caption{Maximal probability of obtaining the inconsistent theory after one update expressed as fraction (\%)
of opinion profiles.}\label{table:max_profiles} \centering \vspace{0.3cm}
\begin{tabular}{cccl}
 \hline\noalign{\smallskip}
  ${}\:M\:$ & ${}\:D\:$ & ${}\:N\:$ & ${}\:F_{\textrm{OP}}$ (\%)\\
 \noalign{\smallskip}\hline\noalign{\smallskip}
  2 & 1 & / & /\\
  2 & 2 & 15 & 6.2930\\
  2 & 3 & 15 & 6.4298\\
  2 & 4 & 6 & 5.5932\\
 \noalign{\smallskip}\hline\noalign{\smallskip}
  3 & 1 & / & / \\
  3 & 2 & 44 & 1.0805\\
  3 & 3 & 37 & 2.5250\\
  3 & 4 & 59 & 6.0837\\
  3 & 5 & 235 & 15.633\\
  3 & 6 & 1780 & 16.867\\
  3 & 7 & 10 & 1.3213\\
  3 & 8 & 8 & 1.3079\\
 \noalign{\smallskip}\hline
\end{tabular}
\end{table}

The general impression of the obtained agent- and opinion-profile-based curves in Figure~\ref{Fig:M2M3Stat} is that they vary smoothly in $N$. It seems that their behavior can be described effectively by an equation that has a substantially simpler form than Equation~(\ref{eq:FractionsShorter}). We tried to fit a continuous function to the discrete curves in Figure~\ref{Fig:M2M3Stat}: the results are presented in the Supporting Information.
\section{Conclusions}\label{sec:Conclusion}
We have presented what can plausibly be regarded as an extension of the HK model, which currently is the most popular model for studying the dynamics of epistemically interacting agents. The extension consisted of equipping the agents with the capability of holding belief states significantly richer than the single beliefs the agents in the HK model have. As we pointed out, and as already followed from earlier work on the discursive dilemma, the extension has a price (apart from the greater mathematical and computational complexity), to wit, updating is not guaranteed to preserve consistency of an agent's belief state. The main goal of this paper was to measure this price by determining the probability that an agent in our model indeed updates to the inconsistent theory. We investigated the effect on this probability of three key parameters: the number of agents in a community; the number of atomic sentences taken into consideration by the agents; and the bound of confidence, determining which agents count as an agent's neighbors.

Taking a social engineering perspective, and based on our results, one can make the following general recommendations for avoiding a zero-update: (i) make (if possible) the number $M$ of atomic sentences large; (ii) avoid (if possible) even-numbered groups of agents; (iii) for a given~$M$, choose (if possible) the number $N$ of agents well below or above the maximum in the curves such as are given in Figure~\ref{Fig:M2M3Stat} for the specific cases of $M=2$ and $M=3$; and (iv) let (if possible) the agents adopt either a very low or a very high bound of confidence, $D$, relative to $2^M$.

As we have seen, apart from the trivial cases with $N=2$ or $M=1$, the probability for an agent (or a population) to reach the inconsistent theory after one update is always non-zero given the update rule we considered. But the good news is that an agent always has a probability $<2\%$ of ending up in the inconsistent belief state. By making either the number of agents or the number of atomic sentences large enough, this probability can be made arbitrarily small. Seen on the scale of the whole population, the probability that one agent ends up with a contradiction is---of course---more complex; however, for sufficiently large populations, that value also decreases with~$N$.

Because the update rule of our model seems to apply naturally to the case of a group of experts participating in a Delphi-study, the recommendations derived from this model may be useful in the design of a Delphi-study in which the experts have to state their preferences in the form of a theory. It may well be that the recommendations made here to lower the probability of inconsistencies differ from those that promote other desired features of communication among agents, such as their ability to converge to the truth. Because the present model is a very simple one, we do not claim to have captured all relevant aspects of opinion-revision of experts in Delphi-studies or of real-life communication of scientists or people in general. However, while a limited number of variables makes it easier to investigate and interpret the outcomes obtained in a model, we intend to investigate in future research the effect of some other parameters that have been held fixed in the present study.

In particular, our current model is void of information, in the sense that the agents have no means of gathering information on the actual the state of the world. In other words, we have not specified which theory of the world corresponds to the actual state of the world. However, recall that one can identify ``It is raining and it is not raining'' as a contradiction, without knowing anything about the weather. Likewise, one can identify the inconsistent theory, without knowing anything about the actual world. Thus, even in the absence of information, social aggregation on logically interrelated issues may yield an outcome, which is undesirable on purely logical grounds. In the current study, we have investigated the occurrence of this undesirable outcome in isolation, without any contingent information present in the system. In future work, we will study agents who can obtain (partial) information about the true theory of the world. The current study on an information-free system will then serve as a benchmark to isolate one aspect of the agents' interaction.

\section*{Acknowledgement}
We are grateful to two anonymous referees for helpful suggestions.

\appendix
\renewcommand{\theequation}{A.\arabic{equation}}
\setcounter{equation}{0}
\section*{Appendix: Derivation of analytical expressions}
This appendix describes how Equation~(\ref{eq:FractionsShorter}) for the agent- and population-based fraction of zero-updates is derived. In the course of this derivation, notions such as `possible world', `opinion profile', and `update rule' which have been introduced in the main text in informal terms receive a formal definition applicable to our model.

First, we take the $M$ atomic sentences to be numbered (arbitrarily) from $0$ to $M-1$. We can thus characterize possible worlds by means of bit-strings of length $M$: if the bit $b_{m}(w)$, with $m \in
\{0,\ldots,M-1\}$, is 1/0, the atomic sentence~$m$ is true/false in world~$w$. There are $w_{\max} = 2^M$
such bit-strings.

These $2^M$ bit-strings of length~$M$ can be numbered as well, most conveniently by their binary value. Each
agent is in a particular belief state, which can be thought of as a set of worlds that the agent deems
possible. Thus, the belief state of an agent can be represented as a longer bit-string of length $w_{\max} =
2^M$ with a single bit for each world equal to 1 or 0 depending on whether the agent deems that world
possible or not. This results in $t_{\max} = 2^{w_{\max}} = 2^{2^M}$ different theories and thus,
correspondingly, possible belief states. Each theory $t$, with $t \in \{0,\ldots,t_{\max}-1\}$, can be
written as $w_{\max}$ bits, $B_{w}(t)$, with $w \in \{0,\ldots,w_{\max}-1\}$, such that:
\begin{equation*}
t \:\: = \:\: \sum_{w=0}^{w_{\max}-1} B_{w}(t)\ 2^{w}.
\end{equation*}
One readily verifies that, given this notation, theory $t=t_{\max}-1$ assigns $1$ to all possible worlds
indeed, and $t=0$ assigns $0$ to all possible worlds.

The sum of bits of a theory $t$---the `bit-sum', written as $s_t$---indicates the number of bits equal to 1
in theory $t$. This can be stated formally as follows:
\begin{equation*}
s_{t} \:\: = \:\: \sum_{w=0}^{w_{\max}-1} B_{w}(t).
\end{equation*}

Recall that, to specify the entire community of all $N$ agents and their belief states, we can count the
number of agents whose belief states are represented by a given theory $t$ and denote it as $n_{t} \in \{0,
\ldots, N \}$, with $\sum_{t=0}^{t_{\max}-1} n_t = N$. Hence, the entire opinion profile is specified
by\cite{fn:AgentTrack}
\begin{equation*}
\overrightarrow{n} \: = \: \langle n_{0}, n_{1}, \ldots, n_{t_{\max}-1}\rangle.
\end{equation*}
Recall further that, because the inconsistent theory $t=0$ is excluded as an initial belief state for all
agents, $n_{0}$ is always~0 in the initial opinion profile.

To determine the new opinion profile of the whole community of agents after one update, we first formalize
how a single `reference' agent updates his belief state, $t_{\mathrm{ref}}$, on the basis of the belief
states of the other agents in the community.

As a first step, the reference agent has to calculate the Hamming distance of the bit-string representing his
belief state, $t_{\mathrm{ref}}$, to that of the bit-strings representing the belief states of the other
agents. The Hamming distance between $t_{\mathrm{ref}}$ and another bit-string $t$ is equal to the bit-sum of
the difference bit-string, which can be found by applying an exclusive-or (XOR) operator: $t\ \textrm{XOR}\
t_{\mathrm{ref}}$. Whereas this operation is familiar in information theory, here we prefer the equivalent
algebraic procedure of performing a bit-wise addition followed by modulo 2:
\begin{equation*}
d(t,t_{\mathrm{ref}}) \:\: = \:\: \sum_{w=0}^{w_{\max}-1} \bigg( \big(B_w(t) + B_w(t_{\mathrm{ref}}) \big)\
\textrm{mod} 2 \bigg).
\end{equation*}

In a community with opinion profile $\overrightarrow{n}$, the number of agents with a belief state at a
distance $d$ from $t_{\mathrm{ref}}$ is called $a_d(t_{ref},\overrightarrow{n})$:
\begin{equation*}
a_d(t_{\mathrm{ref}},\overrightarrow{n}) \:\: = \:\: \sum_{t=0}^{t_{\max}-1} n_t\
\delta_{d,d(t,t_{\mathrm{ref}})},
\end{equation*}
where $\delta_{d,d(t,t_{\mathrm{ref}})}$ is a Dirac delta, which is 0 if the two indices are different and 1
if they are equal.

Now, the reference agent may count the number of agents $A$ within his or her bound of confidence,
$d(t,t_{\mathrm{ref}}) \leqslant D$ (which, it will be recalled, necessarily includes him- or herself) as
follows:
\begin{equation*}
A \:\: = \:\: \sum_{d=0}^D a_d(t_{\mathrm{ref}},\overrightarrow{n}) \:\: = \:\:
a_0(t_{\mathrm{ref}},\overrightarrow{n}) + \ldots + a_D(t_{\mathrm{ref}},\overrightarrow{n}).
\end{equation*}

The next step is to determine how to update any specific bit in the reference agent's belief state,
$B_w(t_{\mathrm{ref}})$. To stay as close as possible to the HK model, the agents in our model take the
arithmetic mean (straight average) over the corresponding bits of the belief states of all their neighbors.
We denote the average using angle brackets:\cite{fn:avgBoC}
\begin{equation}\label{eq:BitAverage}
\big\langle B_w(t_{\mathrm{ref}}) \big\rangle \:\: = \:\: \frac{1}{A} \sum_{d=0}^D \sum_{t=0}^{t_{\max}-1}
B_w(t)\ n_t\ \delta_{d,d(t,t_{\mathrm{ref}})}.
\end{equation}
Based on the above average, the agent decides how to update the value of the $w$\textsuperscript{th} bit. If
the average is smaller than $\frac{1}{2}$, the agent sets this bit to 0; if it is larger than $\frac{1}{2}$,
to 1; and if it is precisely equal to $\frac{1}{2}$, the agent will keep his or her initial value for
that~bit. So, in a sense, we have majority voting here, with the important proviso that the majority is taken
relative to an agent's neighbors and not (necessarily) relative to all agents in his or her community.

The update rule for the $w$\textsuperscript{th} bit of an agent who initially holds theory $t_{\mathrm{ref}}$
is formalized as a function $\textrm{UPD}$:
\begin{eqnarray}\label{eq:UpdateRule}
\lefteqn{\textrm{UPD}\big[ \big\langle B_w(t_{\mathrm{ref}}) \big\rangle \big] \:\: = } \nonumber
\\
&& \left\{ \begin{array}{ll}
1 \quad \quad \quad \quad \quad \ \ \textrm{if} \quad \big\langle B_w(t_{\mathrm{ref}}) \big\rangle > \frac{1}{2} \\[2mm]
B_w(t_{\mathrm{ref}}) \quad \quad \ \ \textrm{if} \quad \big\langle B_w(t_{\mathrm{ref}}) \big\rangle = \frac{1}{2} \\[2mm]
0 \quad \quad \quad \quad \quad \ \ \textrm{if} \quad \big\langle B_w(t_{\mathrm{ref}}) \big\rangle <
\frac{1}{2}.
\end{array} \right.
\end{eqnarray}

To make counting zero-updates more convenient, we introduce a function $\textrm{INV}$ (for `inverse') that
has value 1 if the corresponding updated bit is equal to 0 and vice versa:
\begin{eqnarray*}
\lefteqn{\textrm{INV}\big[ \big\langle B_w(t_{\mathrm{ref}}) \big\rangle \big] \:\: = }
\\
&& \left\{ \begin{array}{ll}
0 \quad \quad \quad \quad \quad \ \ \textrm{if} \quad \big\langle B_w(t_{\mathrm{ref}}) \big\rangle > \frac{1}{2} \\[2mm]
1-B_w(t_{\mathrm{ref}}) \quad \textrm{if} \quad \big\langle B_w(t_{\mathrm{ref}}) \big\rangle = \frac{1}{2} \\[2mm]
1 \quad \quad \quad \quad \quad \ \ \textrm{if} \quad \big\langle B_w(t_{\mathrm{ref}}) \big\rangle <
\frac{1}{2}.
\end{array} \right.
\end{eqnarray*}
An update to the contradiction corresponds to updating all bits to 0. We can count those events by
multiplying the result of $\textrm{INV}$ over all bits: due to the former definition, this product will only
be 1 if all bits are updated to~0.

In order to determine the opinion-profile-based fraction, we introduce a function $\textrm{ZUP}$ (for
`zero-update') that is 1 if there is at least one agent in the community who updates to the contradiction,
and 0 otherwise:
\begin{eqnarray*}
\lefteqn{\textrm{ZUP}(M,N,D,\overrightarrow{n}) \:\: = }
\\
&& \left\{ \begin{array}{ll}
0 \quad \textrm{if} \quad \sum_{t=0}^{t_{\max}-1} \left( \frac{n_{t}}{N} \prod_{w=0}^{w_{\max}-1} \textrm{INV}\big[ \big\langle B_w(t) \big\rangle \big] \right) = 0 \\[2mm]
1 \quad \textrm{if} \quad \sum_{t=0}^{t_{\max}-1} \left( \frac{n_{t}}{N} \prod_{w=0}^{w_{\max}-1}
\textrm{INV}\big[ \big\langle B_w(t) \big\rangle \big] \right) > 0.
\end{array} \right.
\end{eqnarray*}

At this point, we can determine the agent- and opinion-profile-based fractions of zero-up\-dates, $F_{\textrm{AG}}(M,N,D)$ and $F_{\textrm{OP}}(M,N,D)$, by summing over all combinations of the agents' belief
states, $\overrightarrow{n}$. Because each theory $t$, with $t \in \{0,\ldots,t_{\max}-1\}$, is equally
likely to be chosen by all agents, we need to sum over all possible combinations of choices. For the
sum-indices we use the following notation: $t(n)$ is the theory representing agent $n$'s initial belief
state. The requisite functions can then be written as follows:
\begin{subequations}\label{eq:FractionsLong}
\begin{multline}\label{eq:FractionsLongAG}
F_{\textrm{AG}}(M,N,D) \:\:  = \:\: \sum_{t(0)=1}^{t_{\max}-1} \cdots \sum_{t(N-1)=1}^{t_{\max}-1}
\frac{1}{(t_{\max}-1)^N} \:\:\: \times
\\
\qquad\qquad\qquad\quad\: \sum_{t=0}^{t_{\max}-1} \left( \frac{n_{t}}{N} \prod_{w=0}^{w_{\max}-1}
\textrm{INV}\big[ \big\langle B_w(t) \big\rangle \big] \right);
\end{multline}
\begin{multline}\label{eq:FractionsLongOP}
F_{\textrm{OP}}(M,N,D) \:\:  = \:\: \sum_{t(0)=1}^{t_{\max}-1} \cdots \sum_{t(N-1)=1}^{t_{\max}-1}
\frac{1}{(t_{\max}-1)^N} \:\:\: \times
\\
\qquad\qquad\qquad\quad\: \textrm{ZUP}(M,N,D,\overrightarrow{n}).
\end{multline}
\end{subequations}

Because the number of terms is equal to the number of ways the agents can choose a theory as their belief
state, that is \cite{Rosen:2000} (p.~55),
\begin{equation*}
P^{R}(t_{\max}-1,N)=(t_{\max}-1)^N,
\end{equation*}
each term is weighted by the inverse of this.

Equation~(\ref{eq:FractionsLong}) does  not have the exact same form as Equation~(\ref{eq:FractionsShorter}).
To simplify the evaluation of $F_{\textrm{AG}}(M,N,D)$ and $F_{\textrm{OP}}(M,N,D)$, we can reduce the number
of terms drastically by only summing over all \emph{different} anonymous opinion profiles and introducing an
additional weight function (multiset coefficient---see \cite{Rosen:2000}, p.~55) $\frac{N!}{n_{0}!n_{1}!
\cdots n_{t_{\max}-1}!}$:\cite{fn:NoSum}

\begin{multline*}
F_{\textrm{AG}}(M,N,D)\:\: =
\\
\sum_{n_{1}=0}^N \sum_{n_{2}=0}^{N-n_{1}} \cdots
\sum_{n_{t_{\max}-2}=0}^{N-(n_{1}+n_{2}+\cdots+n_{t_{\max}-3})} \frac{N!}{n_{0}!n_{1}! \cdots
n_{t_{\max}-1}!} \:\: \times
\\
\frac{1}{(t_{\max}-1)^N} \sum_{t=0}^{t_{\max}-1} \left( \frac{n_{t}}{N} \prod_{w=0}^{w_{\max}-1}
\textrm{INV}\big[ \big\langle B_w(t) \big\rangle \big] \right);
\end{multline*}
\begin{multline*}
F_{\textrm{OP}}(M,N,D) \:\: =
\\
\sum_{n_{1}=0}^N \sum_{n_{2}=0}^{N-n_{1}} \cdots
\sum_{n_{t_{\max}-2}=0}^{N-(n_{1}+n_{2}+\cdots+n_{t_{\max}-3})} \frac{N!}{n_{0}!n_{1}! \cdots
n_{t_{\max}-1}!} \:\: \times
\\
\frac{1}{(t_{\max}-1)^N} \textrm{ZUP}(M,N,D,\overrightarrow{n}).
\end{multline*}

\noindent This concludes the derivation of Equation~(\ref{eq:FractionsShorter}).

The number of terms in Equation~(\ref{eq:FractionsShorter}) is the multiset coefficient, which represents the
number of ways to choose $N$ out of $t_{\max}-1$ with repetition (see \cite{Rosen:2000} p.~55):
\begin{eqnarray*}
C^{R}(t_{\max}-1,N) & = & \Big( \begin{array}{ccc}
N+t_{\max}-1-1 \\
N
\end{array} \Big)
\\
 & = & \frac{(N+t_{\max}-2)!}{(N)!(t_{\max}-2)!}.
\end{eqnarray*}
This number is smaller than or equal to the number of terms in Equation~(\ref{eq:FractionsLong}),
$P^{R}(t_{\max}-1,N)$.\cite{fn:RedCompCost}

\section*{Supplementary Information: Fitting curves to the statistical data}
\begin{table}[!tb]
\centering \caption{$M=2$. Values of $R^2$ and the three fit parameters with standard error for a log-normal
fit to the agent- and opinion-profile-based $D$-curves, for an even and odd number of
agents.}\label{table:FitM2} \vspace{0.3cm} \scalebox{0.8}{
\begin{tabular}{llll}
 \hline\noalign{\smallskip}
  \multicolumn{4}{c}{$F_{\textrm{AG}}$ at even positions}\\
 \noalign{\smallskip}\hline\noalign{\smallskip}
  $D\:\:\:$ & 2 & 3 & 4\\
 \hline\noalign{\smallskip}
  $R^2$ & 0.946 & 0.9736 & 0.9317\\
  $A$ & 0.0157 $\pm$ 0.0004 & 0.0099 $\pm$ 0.0002 & 0.0152 $\pm$ 0.0004 \\
  $B$ & 0.475\textcolor{white}{0} $\pm$ 0.015 & 0.3654 $\pm$ 0.0094 & 0.511\textcolor{white}{0} $\pm$ 0.017 \\
  $C$ & 9.02\textcolor{white}{00} $\pm$ 0.42 & 7.86\textcolor{white}{00} $\pm$ 0.22 & 9.39\textcolor{white}{00} $\pm$ 0.50\\
 \hline\noalign{\smallskip}
\multicolumn{4}{c}{$F_{\textrm{AG}}$ at odd positions}\\
 \hline\noalign{\smallskip}
  $D$ & 2 & 3 & 4\\
 \hline\noalign{\smallskip}
  $R^2$ & 0.9975 & 0.9988 & 0.9982\\
  $A$ & 0.0102 $\pm$ 7E-5 & 0.0181 $\pm$ 9E-5 & 0.0182 $\pm$ 0.0001 \\
  $B$ & 0.4051 $\pm$ 0.0035 & 0.6744 $\pm$ 0.0047 & 0.6169 $\pm$ 0.0052\\
  $C$ & 6.819\textcolor{white}{0} $\pm$ 0.076 & 4.663\textcolor{white}{0} $\pm$ 0.085 & 4.778\textcolor{white}{0} $\pm$ 0.095 \\
 \hline\noalign{\smallskip}
\multicolumn{4}{c}{$F_{\textrm{OP}}$ at even positions}\\
 \hline\noalign{\smallskip}
  $D$ & 2 & 3 & 4\\
 \hline\noalign{\smallskip}
  $R^2$ & 0.9892 & 0.9516 & 0.9516 \\
  $A$ & 0.0660 $\pm$ 0.0008 & 0.0651 $\pm$ 0.0009 & 0.0500 $\pm$ 0.0015 \\
  $B$ & 0.3342 $\pm$ 0.0050 & 0.577\textcolor{white}{0} $\pm$ 0.014 & 0.423\textcolor{white}{0} $\pm$ 0.014 \\
  $C$ & 14.44\textcolor{white}{0} $\pm$ 0.20 & 16.28\textcolor{white}{0} $\pm$ 0.50 & 8.03\textcolor{white}{00} $\pm$ 0.35\\
 \hline\noalign{\smallskip}
\multicolumn{4}{c}{$F_{\textrm{OP}}$ at odd positions}\\
 \hline\noalign{\smallskip}
  $D$ & 2 & 3 & 4\\
 \hline\noalign{\smallskip}
  $R^2$ & 0.9898 & 0.9934 & 0.9982\\
  $A$ & 0.0663 $\pm$ 0.0008 & 0.0639 $\pm$ 0.0003 & 0.0182 $\pm$ 0.0001 \\
  $B$ & 0.3320 $\pm$ 0.0048 & 0.5444 $\pm$ 0.0045 & 0.6169 $\pm$ 0.0052 \\
  $C$ & 14.51\textcolor{white}{0} $\pm$ 0.20 & 18.24\textcolor{white}{0} $\pm$ 0.18 & 4.778\textcolor{white}{0} $\pm$ 0.095 \\
 \noalign{\smallskip}\hline
\end{tabular}
}
\end{table}

\begin{table*}[!tb]
\centering \caption{$M=3$. Values of $R^2$ and the three fit parameters (with standard error) for a
log-normal fit to the agent- and opinion-profile-based $D$-curves, for an even and odd number of
agents.}\label{table:FitM3} \centering \vspace{0.3cm} \scalebox{0.85}{
\begin{tabular}{llllllll}
 \hline\noalign{\smallskip}
\multicolumn{8}{c}{$F_{\textrm{AG}}$ at even positions}\\
 \noalign{\smallskip}\hline\noalign{\smallskip}
  $D\:\:\:$ & 2 & 3 & 4 & 5 & 6 & 7 & 8\\
 \hline\noalign{\smallskip}
  $R^2$ & 0.9982 & 0.9956 & 0.9831 & 0.9592 & 0.8699 & 0.8747 & 0.8751 \\
  $A$ & 0.0005 $\pm$ 2E-5 & 0.0014 $\pm$ 1E-5 & 0.0024 $\pm$ 3E-5 & 0.0031 $\pm$ 5E-5 & 0.0030 $\pm$ 4E-5 & 0.0030 $\pm$ 4E-5 & 0.0030 $\pm$ 4E-5 \\
  $B$ & 0.3414 $\pm$ 0.0018 & 0.4024 $\pm$ 0.0035 & 0.5294 $\pm$ 0.0096 & 0.749\textcolor{white}{0} $\pm$ 0.025 & 1.190\textcolor{white}{0} $\pm$ 0.046 & 1.172\textcolor{white}{0} $\pm$ 0.045 & 1.170\textcolor{white}{0} $\pm$ 0.045 \\
  $C$ & 21.66\textcolor{white}{0} $\pm$ 0.10 & 13.97\textcolor{white}{0} $\pm$ 0.14 & 15.04\textcolor{white}{0} $\pm$ 0.37 & 25.0\textcolor{white}{0}\textcolor{white}{0} $\pm$ 1.1 & 38.6\textcolor{white}{00} $\pm$ 3.6 & 36.4\textcolor{white}{00} $\pm$ 3.4 & 36.4\textcolor{white}{00} $\pm$ 3.4 \\
 \hline\noalign{\smallskip}
\multicolumn{8}{c}{$F_{\textrm{AG}}$ at odd positions}\\
 \hline\noalign{\smallskip}
  $D$ & 2 & 3 & 4 & 5 & 6 & 7 & 8\\
 \hline\noalign{\smallskip}
  $R^2$ & 0.341 & 0.9973 & 0.9948 & 0.9724 & 0.9891 & 0.9919 & 0.9873 \\
  $A$ & / & 0.0014 $\pm$ 8E-6 & 0.0024 $\pm$ 2E-5 & 0.0031 $\pm$ 4E-5 & 0.0031 $\pm$ 1E-5 & 0.0031 $\pm$ 1E-5 & 0.0031 $\pm$ 1E-5\\
  $B$ & / & 0.4021 $\pm$ 0.0028 & 0.5518 $\pm$ 0.0058 & 0.835\textcolor{white}{0} $\pm$ 0.024 & 1.486\textcolor{white}{0} $\pm$ 0.020 & 1.4734 $\pm$ 0.018 & 1.470\textcolor{white}{0} $\pm$ 0.023 \\
  $C$ & / & 14.05\textcolor{white}{0} $\pm$ 0.11 & 14.21\textcolor{white}{0} $\pm$ 0.21 & 20.71\textcolor{white}{0} $\pm$ 0.95 & 18.47\textcolor{white}{0} $\pm$ 0.84 & 16.84\textcolor{white}{0} $\pm$ 0.69 & 16.91\textcolor{white}{0} $\pm$ 0.87 \\
 \hline\noalign{\smallskip}
\multicolumn{8}{c}{$F_{\textrm{OP}}$ at even positions}\\
 \hline\noalign{\smallskip}
  $D$ & 2 & 3 & 4 & 5 & 6 & 7 & 8\\
 \hline\noalign{\smallskip}
  $R^2$ & 0.9982 & 0.9938 & 0.9888 & 0.9444 & 0.9892 & 0.1241 & 0.8755 \\
  $A$ & 0.0108 $\pm$ 4E-5 & 0.0255 $\pm$ 0.0002 & 0.0601 $\pm$ 0.0005 & 0.1268 $\pm$ 0.0034 & $0.217\mspace{20mu} \pm$ 0.019 & 0.0097 $\pm$ 0.0002 & 0.0114 $\pm$ 0.0003 \\
  $B$ & 0.3105 $\pm$ 0.0018 & 0.3607 $\pm$ 0.0039 & 0.4290 $\pm$ 0.0085 & 0.470\textcolor{white}{0} $\pm$ 0.013 & $1.197\mspace{20mu} \pm$ 0.068 & 2.54\textcolor{white}{00} $\pm$ 0.58 & 0.668\textcolor{white}{0} $\pm$ 0.034 \\
  $C$ & 41.54\textcolor{white}{0} $\pm$ 0.15 & 32.86\textcolor{white}{0} $\pm$ 0.26 & 57.02\textcolor{white}{0} $\pm$ 0.83 & 162.1\textcolor{white}{0} $\pm$ 5.9 & 148E+2 $\pm$ 62E+2 & 44\textcolor{white}{0.00} $\pm$ 29 & 13.0\textcolor{white}{00} $\pm$ 1.0 \\
 \hline\noalign{\smallskip}
\multicolumn{8}{c}{$F_{\textrm{OP}}$ at odd positions}\\
 \hline\noalign{\smallskip}
  $D$ & 2 & 3 & 4 & 5 & 6 & 7 & 8\\
 \hline\noalign{\smallskip}
  $R^2$ & 0.9983 & 0.9933 & 0.9883 & 0.9456 & 0.9917 & 0.8524 & 0.663 \\
  $A$ & 0.0108 $\pm$ 4E-3 & 0.0255 $\pm$ 0.0002 & 0.0601 $\pm$ 0.0005 & 0.1266 $\pm$ 0.0033 & 0.233\textcolor{white}{.0} $\pm$ 0.022 & 0.0112 $\pm$ 0.0002 & / \\
  $B$ & 0.3086 $\pm$ 0.0018 & 0.3607 $\pm$ 0.0040 & 0.4284 $\pm$ 0.0085 & 0.469\textcolor{white}{0} $\pm$ 0.013 & 1.234\textcolor{white}{.0} $\pm$ 0.067 & 1.074\textcolor{white}{0} $\pm$ 0.037 & / \\
  $C$ & 41.58\textcolor{white}{0} $\pm$ 0.14 & 32.93\textcolor{white}{0} $\pm$ 0.27 & 56.96\textcolor{white}{0} $\pm$ 0.83 & 161.3\textcolor{white}{0} $\pm$ 5.8 & 194E+2 $\pm$ 84E+2 & 206\textcolor{white}{.00} $\pm$ 12 & / \\
 \noalign{\smallskip}\hline
\end{tabular}
}
\end{table*}

We tried to fit different asymmetric peak shapes---such as Poisson, Weibull and log-normal---to the data
series presented in Figure~\ref{Fig:M2M3Stat}. The fitting procedure was performed with commercial software
(SigmaPlot). Although no single equation resulted in least-square fits with good $R^2$ values for (almost)
all agent- and opinion-profile-based $D$-curves, a three parameter log-normal distribution gave the best
overall result. Its distribution function is given by:
\begin{equation*}
f(N) = A e^{-\frac{1}{2}\Big(\frac{\log\frac{N}{C}}{B}\Big)^2}
\end{equation*}
The goal of a least-squares fit is to determine the values of the parameters---here $A$, $B$, and $C$---such
that the sum of the squares of the distance between the data points and the value of the fit-curve is
minimal.

To avoid deterioration of the fit quality due to the odd--even wobble, we have split the data sets into
separate files for the values at odd and at even numbers of agents prior to the fitting procedure. The
results for all odd and even, agent- and opinion-profile-based $D$-curves can be found in
Table~\ref{table:FitM2} for $M=2$ and in Table~\ref{table:FitM3} for $M=3$. For $M=2$, $R^2 > 0.9$ for all
curves. For $M=3$, there are eight curves with $R^2 < 0.9$, five of which with $R^2$ between 0.8 and 0.9. For
the three remaining cases with $R^2 < 0.8$ ($F_{\textrm{AG}}$ for $N$ odd and $D=2$, $F_{\textrm{OP}}$ for
$N$ even and $D=7$, and $F_{\textrm{OP}}$ for $N$ odd and $D=8$) the values for the parameters $A$, $B$, and
$C$ are not shown.

\begin{figure}[!tb]
\centering
  \includegraphics[width=8.0cm,keepaspectratio=true]{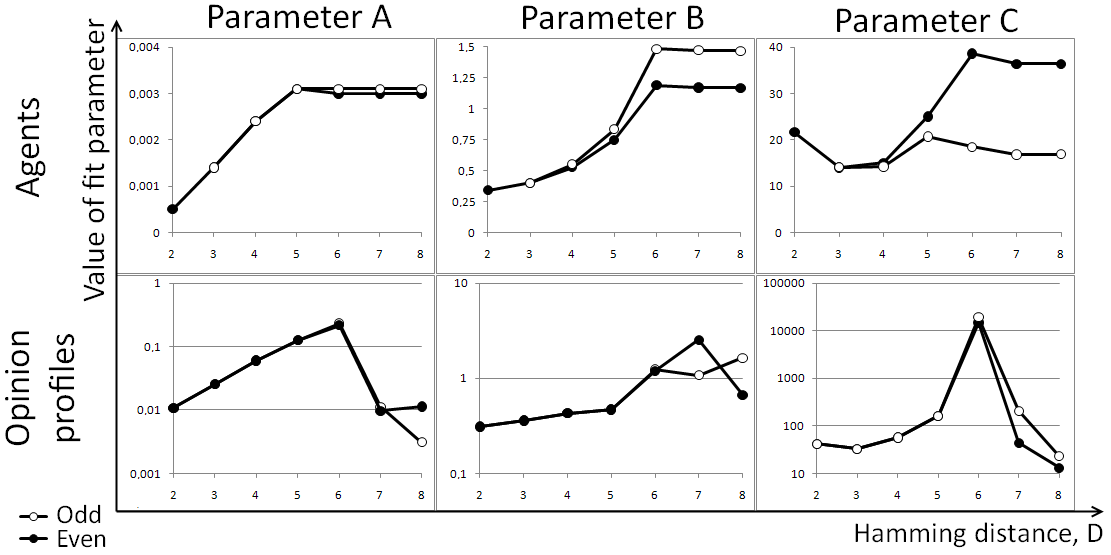}
  \caption{Values of parameters obtained by fitting a log-normal curve to the agent- and opinion-profile-based $D$-curves for the case $M=3$. The filled dots represent the values for an even number of agents, the open dots for odd numbers. For the agent-based values, a linear scale is used; for the opinion-profile-values, a logarithmic scale.}\label{Fig:FitPars}
\end{figure}

A graphical representation of the case where $M=3$ can be found in Figure~\ref{Fig:FitPars}. These curves
suggest a relation among the obtained parameter values, especially for the agent-based curves. For instance,
the fit parameters for the even and odd case correspond well for low $D$-values, but diverge at higher
values, and more drastically so ranging from parameter $A$ over $B$ to $C$. In all curves, the behavior
changes at $D=5$ or $D=6$. For the agent-data, parameter $A$ starts off with a linear trend in $D$, reaching
a plateau from $D=5$ on. For the opinion profile data, the initial trend of parameter $A$ is exponential
linear on the log-scale); this only changes at $D=6$. For parameters $B$ and $C$, the initial behavior in
terms of $D$ increasingly deviates from linearity (or exponential behavior in the opinion profile case).
Though suggestive, the number of data points is insufficient to predict the shape of $D$-curves for higher
values of~$M$.


\bibliography{ProbabilityInconsistencies}

\begin{thebibliography}{53}
\bibitem[{fn:({\natexlab{d}})}]{fn:deftheorem}
\bibinfo{note}{By defining a theory in terms of possible worlds, we have chosen
  for a semantic approach. The equivalent syntactical approach would define a
  theory as a subset of sentences in the agents' language closed under the
  consequence relation for that language.}
\bibitem[{fn:({\natexlab{e}})}]{fn:defudm}
\bibinfo{note}{The precise definition of the update method is stated in the
  appendix. Relative to an agent who initially holds theory $t_{\mathrm{ref}}$,
  Equation~(\ref{eq:BitAverage}) expresses how to calculate the straight
  average of the $w$\textsuperscript{th} bit within the agent's bound of
  confidence. Equation~(\ref{eq:UpdateRule}) formalizes the subsequent rounding
  step of the update rule for the $w$\textsuperscript{th} bit of the agent's
  theory.}
\bibitem[{fn:({\natexlab{g}})}]{fn:M2max}
\bibinfo{note}{For $M=2$, all maxima occur for $N<100$; hence, the data points
  represented in Figure~\ref{Fig:M2M3Stat} suffice. For $M=3$, in all but two
  cases the maximum occurs for $N<110$ and the data used to make
  Figure~\ref{Fig:M2M3Stat} suffice; only for $M=3$ and $D=5$ or $D=6$
  additional data points had to be calculated.}
\bibitem[{fn:({\natexlab{i}})}]{fn:avgBoC}
\bibinfo{note}{Given that the average depends on the bound of confidence $D$
  and the opinion profile of the community $\overrightarrow{n}$, we could put
  those next to the brackets as a subscript. However, to keep the notation
  light, we have omitted this.}
\bibitem[{fn:({\natexlab{j}})}]{fn:NoSum}
\bibinfo{note}{There is no summation over $n_{t_{\max}-1}$, because we know by
  the condition $\sum_{t=0}^{t_{\max}-1} n_t = N$ that $n_{t_{\max}-1} =
  N-(n_{1}+n_{2}+\cdots+n_{t_{\max}-2})$.}
\bibitem[{fn:({\natexlab{k}})}]{fn:RedCompCost}
\bibinfo{note}{To illustrate how drastic the reduction in computational cost
  is, consider the case where $M=2$ and $N=12$:
  Equation~(\ref{eq:FractionsLong}) has $P^{R}(t_{\max}-1,N)$ =
  129~746~337~890~625 terms, whereas Equation~(\ref{eq:FractionsShorter}) has
  `only' $C^{R}(t_{\max}-1,N)$ = 9~657~700 terms. This means that the number of
  evaluations in this example is more than 13 million times smaller for
  Equation~(\ref{eq:FractionsShorter}) as compared to
  Equation~(\ref{eq:FractionsLong}).}
\end{thebibliography}

\end{document}